\pdfoutput=1
\documentclass[sigconf,table,xcdraw,screen,pbalance,authorversion]{acmart}

\usepackage[T1]{fontenc}
\usepackage{multicol}
\usepackage{amsmath,amsfonts,amsthm}
\usepackage{etoolbox}
\usepackage{algorithm}
\usepackage{algpseudocode}
\usepackage{graphicx}
\usepackage{textcomp}
\usepackage{fontawesome5}
\usepackage{booktabs}
\usepackage{makecell}
\usepackage{longtable}
\usepackage{multirow}
\usepackage{float}
\usepackage{multicol}
\usepackage{svg}
\usepackage[frozencache,cachedir=./minted-cache]{minted}
\usepackage{subcaption}
\usepackage{threeparttable}
\usepackage{pifont}
\usepackage{xspace}
\usepackage{tcolorbox}
\tcbuselibrary{breakable, fitting}

\usepackage{enumitem}

\copyrightyear{2024}
\acmYear{2024}
\setcopyright{acmlicensed}
\acmConference[ICSE '24]{2024 IEEE/ACM 46th International Conference on Software Engineering}{April 14--20, 2024}{Lisbon, Portugal}
\acmBooktitle{2024 IEEE/ACM 46th International Conference on Software Engineering (ICSE '24), April 14--20, 2024, Lisbon, Portugal}
\acmDOI{10.1145/3597503.3639231}
\acmISBN{979-8-4007-0217-4/24/04}

\robustify\cite

\newcommand\myding[1]{\ding{\numexpr181+#1\relax}}

\newcommand{\appname}{{\sc Ppt4J}\xspace }
\newcommand\appvar[1]{{\sc Ppt4J}\underline{~~}$\boldsymbol{\Delta}$\textbf{#1}\xspace }
\newcommand{\appfullname}{\textbf{\underline{P}}atch \textbf{\underline{P}}resence \textbf{\underline{T}}est \textbf{\underline{for}} \textbf{\underline{J}}ava Binaries\xspace }

\newcommand\rqsum[2]{
\begin{tcolorbox}[opacityback=1, opacitybacktitle=0, boxrule=1pt]
\textbf{Answer to RQ#1}: #2
\end{tcolorbox}
}

\newcommand\dtname[1]{$\boldsymbol{\mathcal{D}}$\textbf{#1}}

\AtEndPreamble{%
\theoremstyle{acmdefinition}
\newtheorem*{mydef}{Definition}}
\AtBeginEnvironment{mydef}{\setlength{\parskip}{0pt}}

\newcommand\diff{\textit{diff}\xspace}

\makeatletter
\def\input@path{{sections/}}
\makeatother

\makeatletter
\algrenewcommand\ALG@beginalgorithmic{\small}
\makeatother

\graphicspath{figures/}

\begin{document}

\begin{abstract}
The number of vulnerabilities reported in open source software has increased substantially in recent years. Security patches provide the necessary measures to protect software from attacks and vulnerabilities. In practice, it is difficult to identify whether patches have been integrated into software, especially if we only have binary files. Therefore, the ability to test whether a patch is applied to the target binary, a.k.a. patch presence test, is crucial for practitioners. However, it is challenging to obtain accurate semantic information from patches, which could lead to incorrect results.

In this paper, we propose a new patch presence test framework named \appname (\appfullname). \appname is designed for open-source Java libraries. It takes Java binaries (i.e. bytecode files) as input, extracts semantic information from patches, and uses feature-based techniques to identify patch lines in the binaries. To evaluate the effectiveness of our proposed approach \appname, we construct a dataset with binaries that include 110 vulnerabilities.
The results show that \appname achieves an F1 score of 98.5\% with reasonable efficiency, improving the baseline by 14.2\%. 
Furthermore, we conduct an in-the-wild evaluation of \appname on JetBrains IntelliJ IDEA. The results suggest that a third-party library included in the software is not patched for two CVEs, and we have reported this potential security problem to the vendor.
\end{abstract}


\title{\textsc{Ppt4J}: Patch Presence Test for Java Binaries}

\newcommand{\zju}{The State Key Laboratory of Blockchain and Data Security, Zhejiang University}
\author{Zhiyuan Pan}
\affiliation{%
  \institution{\zju}
  \city{Hangzhou}
  \country{China}
}
\email{zy\_pan@zju.edu.cn}

\author{Xing Hu}
\authornote{Corresponding Author}
\affiliation{
  \institution{\zju}
  \city{Hangzhou}
  \country{China}
}
\email{xinghu@zju.edu.cn}
            
\author{Xin Xia}
\affiliation{    
  \institution{Software Engineering Application Technology Lab, Huawei}
  \city{Hangzhou}
  \country{China}
}
\email{xin.xia@acm.org}

\author{Xian Zhan}
\affiliation{
  \institution{The Hong Kong Polytechnic University}
  \city{Hong Kong}
  \country{China}
}
\email{chichoxian@gmail.com}

\author{David Lo}
\affiliation{
  \institution{School of Computing and Information Systems, Singapore Management University}
  \city{}
  \country{Singapore}
}
\email{davidlo@smu.edu.sg}

\author{Xiaohu Yang}
\affiliation{
  \institution{\zju}
  \city{Hangzhou}
  \country{China}
}
\email{yangxh@zju.edu.cn}

\maketitle

\section{Introduction}

The reuse of open source libraries is widespread~\cite{code_reuse}. Vulnerabilities in open-source software have become a major concern, posing significant threats to software and users~\cite{vuln_empirical}. For example, the number of Common Vulnerabilities and Exposures (CVEs~\cite{cve}) reported during the first quarter of 2023 has already exceeded the total number of CVEs reported in 2016~\cite{cve_by_year}. Although upstream developers may discover and fix these vulnerabilities over time, vulnerable versions may still propagate to downstream software or libraries, potentially compromising the security of the systems that rely on them. 

Since open source libraries are frequently distributed as binary files, it is essential that developers and users in the software supply chain be aware of potential vulnerabilities in the libraries they introduce. For example, \textit{Log4Shell} (CVE-2021-44228~\cite{log4shell}) is a well-known vulnerability in Apache Log4j~\cite{log4j} that can lead to the execution of arbitrary code. If a software development team has integrated Log4j into their project, developers should verify if their version of Log4j is vulnerable to \textit{Log4Shell}. In other words, they have to confirm whether the binary contains the patch for CVE-2021-44228.


The above process of testing whether a security patch is applied to the program binaries is called \textit{patch presence test}~\cite{fiber}. However, traditional approaches that only take binary files for analysis cannot be utilized for patch presence tests due to the coarse granularity~\cite{fiber}. For example, binary bug search tools, such as Genius proposed by Feng et al.~\cite{feng2016genius}, identify vulnerability types but cannot test the presence of an arbitrary patch commit. Similarly, binary code search tools, such as Tracy proposed by David et al.~\cite{tracy}, find similar functions but cannot tell whether the function is patched or not.

To accurately test the presence of a patch in fine granularity, Zhang et al. propose FIBER~\cite{fiber}, a patch presence test framework for C/C++ binaries that extracts a localized part of patch \diff~\cite{unixdiff} for signature generation. BScout~\cite{dai2020bscout}, another framework that targets Java binaries, utilizes the entire patch \textit{diff}.


However, existing studies on patch presence test for Java binaries still have the following two limitations:

\begin{itemize}
    \item \textbf{Inability to capture minor changes.} The features extracted by current approaches are unable to handle some subtle modifications to the source code (e.g., changing method call parameters, branch conditions and statements outside method bodies).
    \item \textbf{Limited patch semantic.} The \textit{diffs} utilized in existing approaches cannot perfectly reflect semantic changes (i.e., actual discrepancies in program behavior) in the patches, leading to the inclusion of extraneous information that does not exist in the patches.
\end{itemize}

Given the widespread application of Java, such as server-side programs and Android applications, the limitations mentioned above highlight the need for a more comprehensive and accurate approach. 
To address these limitations, we propose a new patch presence test framework named \appname (\appfullname). 
\appname exploits the correspondence between source code features and binary features. It first extracts features from the source code to generate semantic changes. 
Then, the semantic changes guide \appname to perform source-to-binary feature matching and feature queries. 
Finally, \appname provides the test result by summarizing the queries. 

To evaluate the effectiveness of \appname, we construct a dataset with binaries that include 110 vulnerabilities in total. 
We compare \appname with a state-of-the-art Java patch presence test framework, BScout~\cite{dai2020bscout}, in terms of accuracy, precision, recall, and F1 score. 
To evaluate the effectiveness of \appname in real-world software, we perform an in-the-wild evaluation by testing the presence of patches in various versions of JetBrains IntelliJ IDEA~\cite{intellij_idea}. The results on the dataset demonstrate that \appname achieves an F1 score of 98.5\%, improving the baseline by 14.2\% while maintaining reasonable efficiency. The results of the in-the-wild evaluation indicate that \appname remains accurate in real-world scenarios. In addition, our evaluation reveals that \appname has the ability to uncover instances where application vendors (e.g., JetBrains) have failed to apply security patches to third-party libraries they import, further demonstrating its utility in practical settings.

In summary, we make the following contributions:

\begin{enumerate}
    \item We propose a novel framework for Java binaries, \appname, to accurately test the presence of patches by highlighting semantic code differences from patches.
    \item We construct a dataset to evaluate the effectiveness of \appname by obtaining source code from GitHub repositories, labeling ground truths, and building binaries.
    \item We evaluate \appname using the dataset and real-world software. \appname outperforms the baseline and is able to test the presence of patches in real-world scenarios.
    \item We release the replication package of \appname\footnote{\url{https://github.com/pan2013e/ppt4j}}, including the source code of \appname and our dataset, to facilitate future research.
\end{enumerate}

The remainder of the paper is structured as follows. Section~\ref{sec:bg} introduces the background concepts for this paper and motivates the problem using an example. Section~\ref{sec:approach} presents the approach of \appname. Section~\ref{sec:exp} describes the baseline approach, the preparation for our dataset, and the evaluation metrics. Section~\ref{sec:eval} presents the experimental results and our case study. Section~\ref{sec:disc} and Section~\ref{sec:discussion} discusses comparsions with the baseline and threats of validity. Section~\ref{sec:related} summarizes the related work. Section~\ref{sec:conclusion} concludes the paper.

\section{Background}
\label{sec:bg}
In this section, we discuss the background of the patch presence test task and provide a motivating example to illustrate the main idea of \appname.

\subsection{Patch Presence Test}

The task of patch presence test and its scope of the problem are defined as follows:

\begin{mydef}
Given a security patch \diff $P$ from a specific open source library in the upstream, patch presence test works by evaluating target program binaries $B$ on a boolean function $$ f: (C_1, C_2, P, B) \rightarrow \{\mathrm{True}, \mathrm{False}\}, $$
where $C_1 \text{ and } C_2$ refer to the upstream source code of the software right before and after patch $P$. $B$ refer to the standard Java bytecode in this paper and should be provided by the user. Among $C_1, C_2$ and $P$, at least two of them must be provided and the third can be automatically derived.
\end{mydef}

Patch presence test checks if a specific patch is applied to the provided target binaries~\cite{fiber}. In this task, the upstream source code $C_1, C_2$ and the target binaries $B$ are supposed to belong to the same library, but the target binaries can be compiled from any version of source code. 
The function $f$ has two possible return values. 
If $f$ returns true, we confirm the existence of a patch commit within the binaries, and vice versa.

The advantage of patch presence test lies in its ability to detect specific patch commits. This enables users to specifically check for security patches that they are most concerned about. Thus, the potential risks that arise from vulnerabilities can be mitigated.

\begin{figure}[htbp]
\begin{subfigure}{\columnwidth}
\centering
\begin{minted}[frame=single, fontsize=\scriptsize, breaklines, linenos, escapeinside=||, numbersep=2pt]{diff}
- XmlPullParser        parser = Xml.newPullParser();
- XPPAttributesWrapper attributes = new XPPAttributesWrapper(parser);
+ try
+ {
+   XmlPullParser        parser = Xml.newPullParser();
+   XPPAttributesWrapper attributes = new XPPAttributesWrapper(parser);
\end{minted}
\caption{CVE-2017-1000498~\cite{CVE-2017-1000498} }
\label{fig:diff_case1}
\end{subfigure}

\begin{subfigure}{\columnwidth}
\centering
\begin{minted}[frame=single, fontsize=\scriptsize, breaklines, linenos, escapeinside=||, numbersep=2pt, breakanywhere]{diff}
- Document<T> doc = parser.parse(is);
+ XMLStreamReader reader = StaxUtils.createXMLStreamReader(is);
+ Document<T> doc = parser.parse(reader);
\end{minted}
\label{fig:diff2}
\caption{CVE-2016-8739~\cite{CVE-2016-8739} }
\label{fig:diff_case2}
\end{subfigure}

\begin{subfigure}{\columnwidth}
\centering
\begin{minted}[frame=single, fontsize=\scriptsize, breaklines, linenos, escapeinside=||, numbersep=2pt]{diff}
-    if (A == Algorithm.none && B == 2 && C == 0) {
-      return Mapper.deserialize(base64Decode(...), JWT.class);
+    if (B == 2 && C == 0) {
+      if (A == Algorithm.none) {
+        return Mapper.deserialize(base64Decode(...), JWT.class);
+      } else {
+        throw new InvalidJWTSignatureException();
+      }
\end{minted}
\caption{CVE-2018-11797~\cite{CVE-2018-11797} }
\label{fig:diff_case3}
\end{subfigure}
\Description{Examples of ``semantic redundancy'': \diff lines excerpted from full security patches.}
\caption{Examples of ``semantic redundancy'': \diff lines excerpted from full security patches.}
\label{fig:diff_cases}
\end{figure}

\begin{figure*}[htbp]
    \centering
    \includegraphics[width=0.95\linewidth]{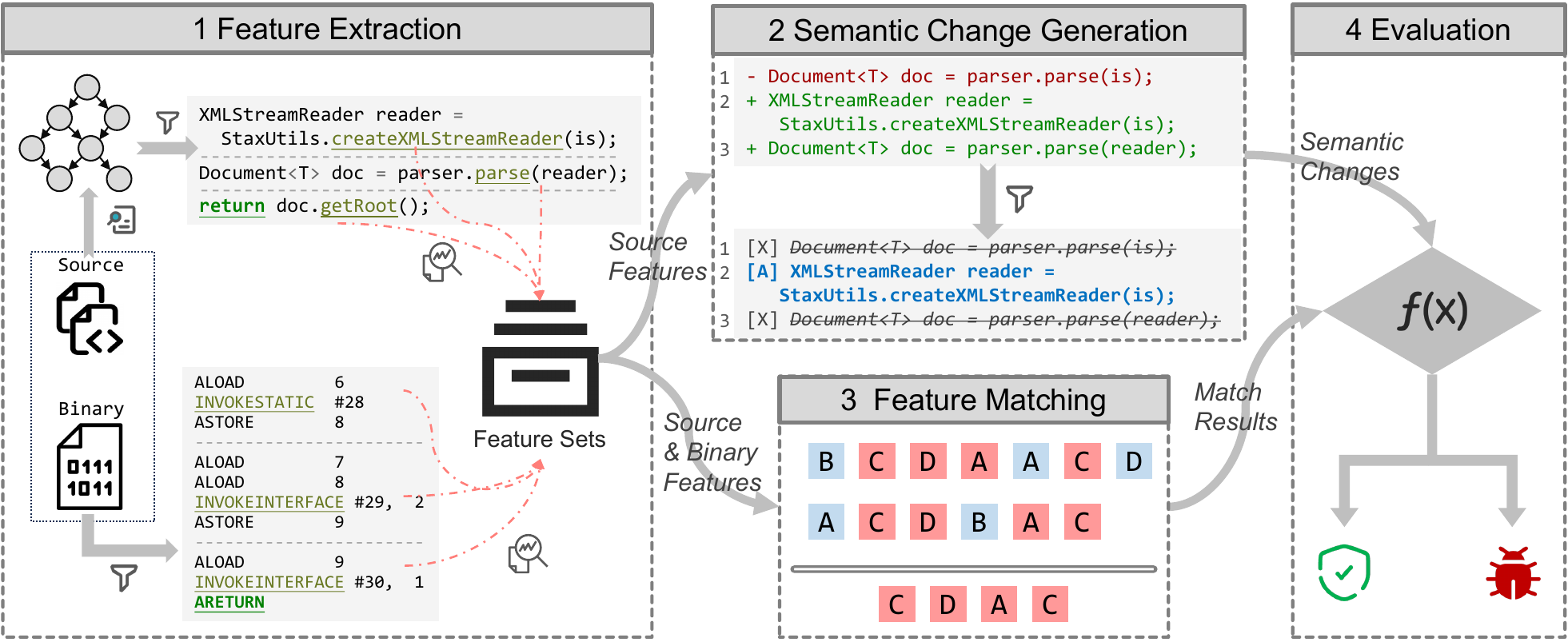}
    \Description{Overall approach of \textsc{Ppt4J}}
    \caption{Overall approach of \textsc{Ppt4J}}
    \label{fig:arch}
\end{figure*}

\subsection{Motivating Example}
\label{sec:motivation}

Existing patch presence test work for Java binaries takes the complete \diff (i.e., the exact differences in characters) as input to analyze patch presence. 
However, existing work is limited as the whole \diff will introduce ``semantic redundancy'', that is, some \verb|(-)| and \verb|(+)| lines end up with no semantic changes, which may introduce unrelated information.

Intuitively, to accurately test the presence of a patch, we should extract precise semantic changes from \diff that reflect all semantic information while not including unrelated information. We illustrate the importance of semantic changes by using real-world examples shown in Figure~\ref{fig:diff_cases}.

In the first case of Figure~\ref{fig:diff_cases}, the \diff introduces a \verb|try-catch| block with six \diff lines. 
However, there are no differences between lines 1, 2, and 5, 6, except for indentations. If we trim the leading spaces/tabs, these lines would be identical. In other words, the actual semantic-related change of the patch is the \verb|try-catch| environment, rather than these four statements. Similar cases include moving statements into \verb|if-else|, \verb|for|, or \verb|while| blocks. In the second case (Figure~\ref{fig:diff_case2}), lines 1 and 3 differ due to the change of a variable (i.e., \verb|is| $\longrightarrow$ \verb|reader|). The change originates from calling the method \verb|createXMLStreamReader|. Thus, the code that causes the actual semantic change is line 2, not lines 1 and 3. In the third case (Figure~\ref{fig:diff_case3})\footnote{The code in the figure is simplified for illustration. The original version is available at \url{https://github.com/FusionAuth/fusionauth-jwt}, with commit hash 0d94dcef0133d699f21d217e922564adbb83a227}, we can observe that an \verb|if| block is deleted and another \verb|if| block is added. But after inspecting the code, the only semantic change is line 7, that is, throw an exception when \verb|B == 2 && C == 0 && A != Algorithm.none|.

From these examples, we conclude that the differences in characters may not correspond to semantic changes. Motivated by this, we illustrate the idea of our work, which highlights semantic changes and is carried out in the following steps:

\begin{enumerate}
\renewcommand{\labelenumi}{Step~\theenumi}
    \item \textbf{Feature extraction.} We extract features for patch-relevant Java source code and binaries. 
    \item \textbf{Semantic change generation.} We generate semantic changes for a patch based on the original \diff and the extracted features in Step~1.
    \item \textbf{Feature matching.} We match sequences of source-level features and binary-level features extracted in Step~1.
    \item \textbf{Patch presence evaluation.} Instructed by semantic changes in Step~2 and feature matching results in Step~3, we evaluated the patch presence for the binaries.
\end{enumerate}

\section{Approach}
\label{sec:approach}

In this section, we first provide an overview of \appname and its architecture. Then, we introduce each component in detail.

\subsection{Overview of \textsc{Ppt4J}}

The framework of \appname is depicted in Figure~\ref{fig:arch}. \appname takes the source code and binaries as input. The output of the framework is the patch presence status (i.e., \verb|true| or \verb|false|) of the binaries. Additionally, the framework provides user-configurable parameters, which we will discuss in detail later in this section. By utilizing this framework, \appname is able to efficiently analyze binaries and accurately test the presence of patches.

\subsection{Feature Extraction}
\label{sec:feat}

This component extracts rule-based features for Java code lines and bytecode blocks.

\subsubsection{Pre-Process}

Raw source code and binaries are not ideal for feature extraction due to certain programming practices, e.g., a statement might be splitted into multiple lines. 
Additionally, a single instruction may not accurately represent the intended semantic information of a Java statement, which is usually compiled into a group of instructions.

To address this issue, \appname incorporates filters that aggregate discrete elements. For instance, we merge split lines in Java source code to create logical lines from abstract syntax trees (ASTs). 
As for binaries, we propose to split instruction sequences into logical blocks using line number information in binaries. Sometimes, line numbers may be absent or stripped. We discuss such case in Section~\ref{sec:line_number}.

\subsubsection{Feature Types}
\label{sec:feat_types}

\begin{figure}[htbp]
\centering
\includegraphics[width=0.8\columnwidth]{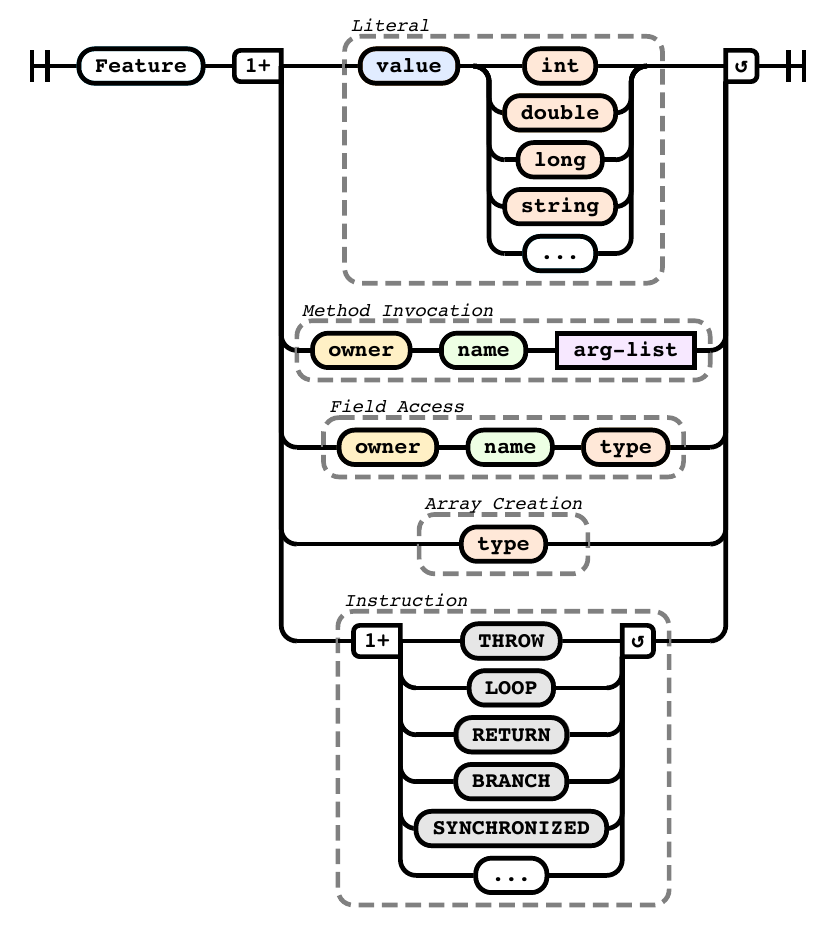}
\Description{The railroad diagram that illustrates feature types selected in our approach.}
\caption{The railroad diagram that illustrates feature types selected in our approach.  }
\label{fig:feat_def}
\end{figure}

Each line of logical source code or each logical bytecode block corresponds to a set of features that can include zero or more features. Our goal is to select feature types that reflect a large proportion of the Java language specifications and Java VM specifications~\cite{java_se}. To achieve this, we select a variety of simple and non-trivial feature types. This ensures that \appname is capable of capturing significant information and minimizing the risk of missing important details. As shown in Figure~\ref{fig:feat_def}, the feature types in our definition include literals, method invocations, field accesses, array creations, and special instructions. 

To be specific, ``literal'' contains compile-time constant values, fields, and expressions. ``Method invocation" contains calls to static methods, virtual methods, and interface methods. We exclude method invocations that can be implicitly generated by the compiler frequently. The excluded methods are \texttt{<init>} in \texttt{Object} and \texttt{String-\\Builder}, \texttt{toString}, \texttt{valueOf}, \texttt{append} and \texttt{longValue}. For other method invocations, we include a method's name, owner (i.e., the class to which it belongs), and actual parameter types. ``Field access'' and ``array creation'', as their names suggest, contain read/write to mutable fields and creation of array objects, respectively. Finally, when extracting ``special instructions'', we seek source code elements or instructions that meet one of the following characteristics:

\begin{enumerate}
    \item Distinctive calculations. \myding{1} Special binary operators: shift and \verb|instanceof|; \myding{2} Special unary operators: \texttt{++} and \texttt{-{}-} (including prefix and postfix).
    \item Control flow manipulations: \verb|return| statements, \verb|throw| statements, \verb|if| blocks and loops.
    \item Synchronization primitives, e.g., \verb|monitorexit| instruction and \verb|synchronized| blocks.
    \item Representations of syntatic sugars, e.g., the labels of \verb|switch| blocks, lambda expressions. 
\end{enumerate}

\subsubsection{Extraction}
\label{sec:feat_ext}

For Java source code, we extract features by analyzing ASTs. 
First, we recursively generate the class dependencies from the \verb|import| statements. Then, we design a custom AST visitor based on Spoon~\cite{spoon}. Specifically, the visitor walks through the syntax tree, from the root node to leaf nodes. If a source code element meets any of the rules described in Section~\ref{sec:feat_types}, it extracts the related values or literals, constructs a feature object, and adds it to the corresponding feature set.
For Java binaries, we extract features by traversing logical blocks using ASM~\cite{asm_website, asm_paper1, asm_paper2}. 
Specifically, for each logical bytecode block, we traverse the instructions in the block. Similar to the extraction of source code, if an instruction meets any of the rules, the related values or literals are used to construct a feature object.

For non-trivial feature types, we perform static analysis to collect fine-grained features, especially when dealing with minor patches. The details can be found as follows:

\myding{1} \textbf{Literals.} To deal with compiler optimizations on literals, we implement constant propagation and folding for the source code. 
Accesses to constant fields are replaced by literals, and arithmetics on literals or constant fields are simplified. This transformation is interprocedural and interclass because ASTs of all dependent classes are accessible. 

\myding{2} \textbf{Method invocations.} Java is an object-oriented programming language, and inheritance is quite common for code reusability. In this case, objects of a derived class can be referenced by variables of a base class. Thus, normal method signatures cannot always reflect the exact types of the arguments of method invocations. To obtain the exact argument types of method calls and build fine-grained method signatures, we implement type analysis for binaries. Starting from the point where an argument variable is initialized, the analysis simulates possible execution paths and returns a more precise type of the variable. To be specific, when walking through a method, it maintains a stack and a local variable table, which store the types of elements\footnote{If an element is null, we denote the type as null and do not step further.}. During this process, it takes out variable types in the stack and local variable table, simulates a bytecode instruction, stores the resulting types back to the data structures, and then moves to the next instruction. We use the common superclass if an element type has multiple possibilities. Due to the static type system in Java, it is ensured that the stored type is a subtype of the type found in the method's original signature. Thus, the type analysis results in a more detailed method invocation feature.

\begin{figure}[htbp]
\centering
\begin{minted}[frame=single, fontsize=\footnotesize, breaklines, linenos, escapeinside=||, numbersep=2pt]{diff}
@@ -400,7 +400,7 @@ private static int fastRound(final float x) {
     private int extend(int v, final int t) {
         // "EXTEND", section F.2.2.1, figure F.12, page 105 of T.81
         int vt = (1 << (t - 1));
-        while (v < vt) {   
+        if (v < vt) {
             vt = (-1 << t) + 1;
             v += vt;
         }
 
\end{minted}
\Description{Patch snippet from CVE-2018-17202~\cite{CVE-2018-17202}}
\caption{Patch snippet from CVE-2018-17202~\cite{CVE-2018-17202}}
\label{fig:if_while}
\end{figure}

\myding{3} \textbf{Loop statements.} We distinguish condition statements and loop statements by generating intra-procedure control flow graphs (CFGs) for binaries and analyzing the graphs. For example, the patch in Figure~\ref{fig:if_while} changes a \verb|while| loop to an \verb|if| statement. Although both of \verb|while| loop and \verb|if| statement generate branch instructions, loop statements generate \verb|GOTO| instructions in future blocks that redirect the control flow. Since the Java programming language does not allow programmers to manipulate control flows with keywords like \verb|goto|, we assert that a backward \verb|GOTO| instruction must point to a loop condition.

\subsection{Semantic Change Generation}
\label{sec:diffgen}

As discussed in Section~\ref{sec:motivation}, commonly used utilities like UNIX \diff compare texts between different versions. However, \diff is limited to differences in characters and cannot explain the behavior of the program~\cite{susan_diff}. To evaluate patch presence more accurately, we extract semantic changes for each security patch based on the original \diff generated from the \texttt{git diff <commit> <commit>} command. This command compares two commits of a git repository and generates output in \textit{unified diff format}~\cite{unified_diff}.

First, we parse the \diff and split the hunks (i.e., groups of differing lines interspersed in files) into finer-grained code blocks. During the parsing process, we filter out unrelated \diff contents, i.e., lines that contain no features. 
Also, split lines are treated as a single logical line. After that, we categorize the blocks into three types, \textit{pure addition} (\textbf{\textit{A}}-type), \textit{pure deletion} (\textbf{\textit{D}}-type), and \textit{mixed blocks} (\textbf{\textit{M}}-type). An \textbf{\textit{M}}-type block consists of both \verb|(+)| lines and \verb|(-)| lines, in which two kinds of lines overlap or adjoin in space, and \verb|(-)| lines appear before \verb|(+)| lines.

Then, we classify the \diff lines as pure \verb|(+)| lines, pure \verb|(-)| lines and \{\verb|(-)| line, \verb|(+)| line\} pairs, and put them in different sets. 
The classification result reflects the semantic of the patch and can be used to instruct the patch presence evaluator (in Section~\ref{sec:patch_eval}). 
We define four types of sets: \textbf{\texttt{SA}} for pure \verb|(+)| lines, \textbf{\texttt{SD}} for pure \verb|(-)| lines, \textbf{\texttt{SX}} for lines to be excluded in our approach, and \textbf{\texttt{SM}} for \{\verb|(-)| line, \verb|(+)| line\} pairs. In particular, \textbf{\texttt{SX}} refers to the lines that do not participate in future evaluations. 
The line pairs in the \textbf{\texttt{SM}} set are called ``modification line pairs''. We consider all \diff lines as semantic changes for \textbf{\textit{A}}-type and \textbf{\textit{D}}-type blocks and put them into \textbf{\texttt{SA}} or \textbf{\texttt{SD}} sets. 
We apply a heuristic approach for \textbf{\textit{M}}-type blocks using previously extracted features to filter semantic changes, as described in Algorithm~\ref{alg:modblk}. The above four sets contain semantic change lines and are utilized in the patch presence evaluation (Step 4 in Figure~\ref{fig:arch}). 

\begin{algorithm}
\caption{Filtering semantic changes for \textbf{\textit{M}}-type blocks}
\label{alg:modblk}
\begin{algorithmic}[1]
\Require An \textbf{\textit{M}}-type diff block \verb|B| \vskip 0.2em
\State \verb|del, add| $\leftarrow$ \verb|splitByType(B)|
\State \verb|c1| $\leftarrow$ \verb|getSmallerOne(del, add)|
\State \verb|c2| $\leftarrow$ \verb|getLargerOne(del, add)|
\State \verb|window_size| $\leftarrow$ \verb|c1.size()|
\State \verb|initWindow(c2, window_size)|\Comment{\textcolor{gray}{Initialize a sliding window on \texttt{c2}}}
\ForAll{possible windows}
\State \verb|best_window| $\leftarrow$ the most similar\footnotemark window to \verb|c1|.
\State \verb|coeff| $\leftarrow$ \verb|simMetric(c1, best_window)|
\EndFor
\If{\texttt{coeff} $= 1$} \Comment{\textcolor{gray}{These lines should be excluded}}
\State \verb|SX.putAll(c1)| 
\State \verb|SX.putAll(best_window)|
\ElsIf{\texttt{coeff} $\ge \sigma_f$} \Comment{\textcolor{gray}{Considered as modification pairs}}
\State \verb|makePairs(c1, best_window).forEach(SM::put)|
\EndIf
\ForAll{remaining line \texttt{l}}
\If{\texttt{l.type} $= (+)$}
\State \verb|SA.put(l)| \Comment{\textcolor{gray}{Considered as addition lines}}
\Else
\State \verb|SD.put(l)| \Comment{\textcolor{gray}{Considered as deletion lines}}
\EndIf
\EndFor
\end{algorithmic}
\end{algorithm}

\footnotetext{Similarity metric for feature sets and threshold parameter $\sigma_f$ are discussed in Section~\ref{sec:feat_match}}

To be specific, Step 2 in Figure~\ref{fig:arch} illustrates an example of semantic change generation. 
In this example, we can observe that lines 1 and 3 do not have semantic changes if we put them together, and thus should be excluded. Line 2, on the other hand, has a semantic change (i.e., function call \verb|createXMLStreamReader|), thus we put it into the \textbf{\texttt{SA}} set.

Our semantic change generator can handle all code differences inside method bodies. However, for out-of-method source code lines, we only consider assignments to fields. Since these assignments are components of the constructor methods \verb|<init>| or static initialization blocks \verb|<clinit>|, which cannot be ignored. 
In this paper, we ignore other types of out-of-method code differences, such as modifying method signatures or implementing new interfaces and we assume that these changes should be reflected by other changes inside the method bodies.

\subsection{Feature Matching}
\label{sec:feat_match}

From the last step, we extract unified features from the source code and the binaries, they can be considered as sorted sequences of feature sets, i.e.,
$$ [(idx_1, S_1), (idx_2, S_2), (idx_3, S_3) \cdots] $$
where $idx_1 < idx_2 < \cdots < idx_n$.

$idx_i$ refers to a logical line number in the source feature sequence or an instruction block index in the binary feature sequence, and
$S_i$ refers to a feature set. For source sequences, $idx$ may not be consecutive due to empty or comment lines. In practice, we remap these indices for convenience.

Inspired by sequence alignment techniques used in other disciplines, we employ a similar approach to match source code features and binary features. In general, given two feature set sequences,
\begin{equation*}
\begin{aligned}
A & = [(a_1, S_{a_1}), \cdots , (a_n, S_{a_n})], \\
B & = [(b_1, S_{b_1}), \cdots , (b_n, S_{b_n})],
\end{aligned}
\end{equation*}
the feature matcher outputs a sequence
$$ C = [(a_{i_1}, b_{j_1}, S_{k_1}), \cdots, (a_{i_n}, b_{j_n}, S_{k_n})]$$
where the first key in the tuple sorts elements. The $a_{i_n}$-th logical source code line matches the $b_{j_n}$-th binary block.
Before matching, we first define the equivalence of two feature sets:


\begin{mydef}
Given two finite feature sets $S_1, S_2$ and a similarity metric $f: (S_1, S_2) \rightarrow [0,1] $,  $S_1$ and $S_2$ are equivalent iff $$ f(S_1, S_2) \ge \sigma_f, \sigma_f \in (0, 1)$$ where $\sigma_f$ is a threshold parameter. By default, $\sigma_f$ is set to 0.7.
\end{mydef}

Specifically, we use the Jaccard similarity coefficient~\cite{jaccard} $\mathcal{J}$ as the similarity metric, that is
\begin{equation*}
\mathcal{J}(A, B)=
\begin{cases}
1 & \text{A and B are both empty} \\
\dfrac{|A\cap B|}{|A \cup B|}  & \text{otherwise}
\end{cases}
\end{equation*}

After defining the equivalence of feature sets, we apply a sequence-matching algorithm. Specifically, we use the longest common subsequence (LCS) algorithm because we should obtain as many matches as possible and we expect the matches to be in order. For example, Step~3 in Figure~\ref{fig:arch} illustrates the matching process of two sequences \verb|BCDAACD| and \verb|ACDBAC|. After applying the LCS algorithm, we get the matching result \verb|CDAC|. In our LCS algorithm, the elements in the sequences are feature sets.
Then, the aforementioned similarity metric and equivalence of two feature sets are applied.

However, due to line breaks in the source code, some source feature sets in the sequence are not matched with binary feature sets.
To address this issue, after running the LCS algorithm, we start a second matching pass for unmatched binary blocks and employ a heuristic approach that attempts to search for the best match. The heuristic idea is that the union of a few consecutive unmatched binary feature sets is likely to match a source feature set. Specifically, we employ a variable-length sliding window to scan the spaces of the unmatched binary feature sets. As the sliding window moves, in each move we obtain an \textit{aggregated feature set}, which simply means the union of all binary feature sets in a window.
If an aggregated feature set corresponds to a source code line, we regard it as a new match. 
Since the scanning process is time-consuming and memory-consuming and we assume that programmers generally avoid splitting statements into too many lines, we limit the maximum window size to increase performance. By default, the size is set to 5 instructions/window.

\subsection{Patch Presence Evaluation}
\label{sec:patch_eval}

The patch presence evaluator serves as the decision-making component in our approach, responsible for aggregating all relevant information and determining the presence of a patch in the binaries. The idea is to assign votes to each semantic change line and let them make recommendations. These semantic change lines are obtained from the output of the semantic change generator as described in Section~\ref{sec:diffgen}.
Our voting rules for the patch presence evaluator are described below: 

\myding{1} There are two voting options: \textbf{A} for those in favor of the binary being \textbf{patched}, and \textbf{B} for those in favor of the binary being \textbf{unpatched}. 

\myding{2} For the \verb|(+)| lines in the \textbf{\texttt{SA}} set, if the lines appear in the binaries, they vote for \textbf{A}. If not, they vote for \textbf{B}. 

\myding{3} Similarly, for the \verb|(-)| lines in the \textbf{\texttt{SD}} set, if the lines \textbf{do not} appear in the binaries, they vote for \textbf{A}. Otherwise, they vote for \textbf{B}.

\myding{4} For modification line pairs \verb|(pre, post)| in the \textbf{\texttt{SM}} set, we expect the features in binaries to be more similar to the post-patch line than to the pre-patch line. Thus, if the similarity score of (\texttt{post}, binary) is larger than (\texttt{pre}, binary), the lines vote for \textbf{A}. Otherwise, they vote for \textbf{B}. In terms of similarity, we use the same metric described in Section~\ref{sec:feat_match}.

\myding{5} Lines in the \textbf{\texttt{SX}} set are ignored as they do not contain semantic changes. 

To determine whether a given source code line appears in the binaries, we check the existence of its corresponding logical line number in the output sequence generated by the feature matcher.
In addition to these rules, lines with a greater number of features are preferred in our approach. To be specific, the number of votes for each line is equivalent to the number of features it possesses.
Once we have all the votes, we calculate the patch's ``support rate'', which refers to the percentage of total votes in favor of the binary being patched. We consider the patch to be present if the support rate equals or exceeds a specified threshold parameter, denoted by $\sigma_p$. By default, the parameter is set to 0.6.

\begin{table*}[htbp]
\caption{List of vulnerabilities selected from Vul4J~\cite{vul4j2022} }
\label{tab:vul4j_cves}
\resizebox{0.9\linewidth}{!}{
\begin{tabular}{@{}ll@{ }||ll@{}}
\toprule
\textbf{Library Name}                          & \textbf{Vulnerability ID}                   & \textbf{Library Name}                          & \textbf{Vulnerability ID}  \\ 
\midrule
alibaba/fastjson                      & CVE-2017-18349                     & eclipse/rdf4j                         & CVE-2018-20227       \\
apache/camel                          & CVE-2015-0264, CVE-2015-0263       & ESAPI/esapi-java-legacy               & CVE-2013-5960                      \\
apache/commons-compress               & CVE-2019-12402, APACHE-COMMONS-001 & esigate/esigate                       & CVE-2018-1000854                   \\
apache/commons-configuration          & CVE-2020-1953                      & FasterXML/jackson-dataformat-xml      & CVE-2016-7051, CVE-2016-3720       \\
apache/commons-fileupload             & CVE-2013-2186                      & inversoft/prime-jwt                   & 
    CVE-2018-1000125, CVE-2018-1000531 \\
apache/commons-imaging                & CVE-2018-17201, CVE-2018-17202     & javamelody/javamelody                 & CVE-2013-4378                      \\
apache/commons-io                     & CVE-2021-29425                     & jenkinsci/ccm-plugin                  & CVE-2018-1000054                   \\
apache/cxf                            & CVE-2015-5253, CVE-2016-8739       & jenkinsci/groovy-sandbox              & CVE-2018-1000865                   \\
apache/httpcomponents-client          & HTTPCLIENT-1803                    & jenkinsci/jenkins                     & CVE-2017-1000355, CVE-2018-1000864, CVE-2018-1999044 \\
apache/jspwiki                        & CVE-2019-0225                      & jenkinsci/junit-plugin                & CVE-2018-1000056                   \\
apache/pdfbox                         & PDFBOX-3341, CVE-2018-11797        & jenkinsci/pipeline-build-step-plugin  & CVE-2018-1000089                   \\
apache/santuario-java                 & CVE-2014-8152                      & jenkinsci/subversion-plugin           & CVE-2018-1000111                   \\
apache/shiro                          & CVE-2016-6802                      & neo4j-contrib/neo4j-apoc-procedures   & CVE-2018-1000820                   \\
apache/sling                          & CVE-2016-5394, CVE-2016-6798       & OpenRefine/OpenRefine                 & CVE-2018-20157, CVE-2018-19859     \\
apache/sling-org-apache-sling-xss     & CVE-2017-15717                     & resteasy/Resteasy                     & CVE-2020-1695                      \\
apache/struts &
  \makecell[tl]{CVE-2016-8738, CVE-2014-0113, CVE-2014-0112, \\ CVE-2016-4465, CVE-2016-2162, CVE-2014-7809, \\ CVE-2016-4436, CVE-2014-0116, CVE-2015-1831, \\ CVE-2016-3081} & spring-projects/spring-data-rest      & CVE-2017-8046            \\
apache/tika                           & CVE-2018-8017                      & spring-projects/spring-framework      & CVE-2018-15756, CVE-2016-9878, CVE-2018-1272 \\
apache/tomee                          & CVE-2015-8581                      & spring-projects/spring-security-oauth & CVE-2016-4977                      \\
apereo/java-cas-client                & CVE-2014-4172                      & square/retrofit                       & CVE-2018-1000850                   \\
cloudfoundry/uaa                      & CVE-2019-3775, CVE-2018-1192       & swagger-api/swagger-parser            & CVE-2017-1000207                   \\
codehaus-plexus/plexus-archiver       & CVE-2018-1002200                   & x-stream/xstream                      & CVE-2019-10173                     \\
(Continued in the right column) &  & zeroturnaround/zt-zip                 & CVE-2018-1002201                   \\
\bottomrule
\end{tabular}
} 
\end{table*}

\section{Experimental Settings}
\label{sec:exp}

In this section, we describe the details of the baseline, dataset, and implementation of our approach.

\subsection{Baseline}

There exist a number of patch presence test approaches~\cite{fiber, pdiff, dai2020bscout, osprey}. However, most of them target C/C++ binaries and cannot be directly employed to test Java binaries because machine instruction sets and bytecode instruction sets differ a lot. Among the existing approaches, BScout~\cite{dai2020bscout} is a patch presence test framework specifically designed for Java binaries, and we use it as our baseline. Since the official implementation of BScout is not publicly available, we reimplement their approach with about 5,000 lines of Java code.

\subsection{Dataset}
\label{sec:dataset}

\begin{figure}[htbp]
\centering
\includegraphics[width=\columnwidth]{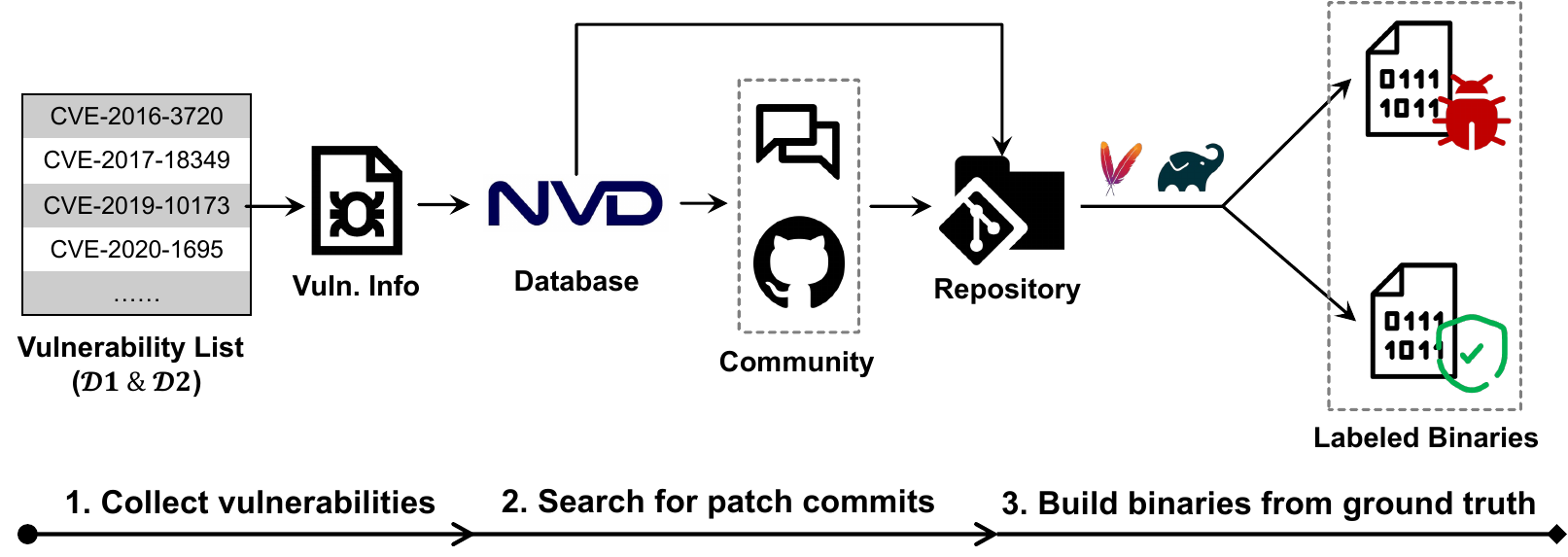}
\Description{Steps to construct the dataset}
\caption{Steps to construct the dataset}
\label{fig:cve_prepare}
\end{figure}

To evaluate the effectiveness of \appname, we construct a dataset, as illustrated in Figure~\ref{fig:cve_prepare}.  In detail, the construction process can be divided into the following three steps:

\myding{1} \textbf{Vulnerability collection.} 
In this step, we collect a number of vulnerabilities for evaluation. We include two lists of vulnerabilities. The first list (abbreviated as \dtname{1}) consists of the Java library vulnerabilities evaluated in BScout's experiments, and the second one (abbreviated as \dtname{2}) is selected\footnote{Some vulnerabilities are excluded because the corresponding projects are too old and the build tools fail to resolve some dependencies during compilation.} from Bui et al.'s work, i.e., Vul4J~\cite{vul4j2022}. The corresponding patches in \dtname{1} are trivial and are used to check the correctness of our reimplementation of BScout. 
To evaluate the ability to test the presence of patches that introduce minor changes, we also include the \dtname{2} list. Detailed vulnerabilities in \dtname{2} are listed in Table~\ref{tab:vul4j_cves}. 

\myding{2} \textbf{Commit searching.} 
In addition to the list of vulnerabilities, the dataset also contains project snapshots and corresponding binaries labeled with ground truth.
For a vulnerability in the list, we first search it in the National Vulnerability Database (NVD)~\cite{nvd}, a CVE database managed by the U.S. government. If a patch commit URL is given in the database, we are done with the search. Otherwise, we manually search for patch-related commit IDs or issue IDs in open-source projects. Once we locate the URL, we can derive the patch status (i.e., patched or unpatched) of a project snapshot after searching. We use patch status as ground truth to label a project snapshot. 


\myding{3} \textbf{Binary compilation.} 
In the final step, we compile binaries from project snapshots. The compilation process can be automated as most open-source Java projects are well documented and provide build scripts , e.g., \verb|build.xml|~\cite{apache_ant}, \verb|pom.xml|~\cite{maven}, \verb|build.gradle|~\cite{gradle} and shell scripts. In practice, most of the projects in the dataset~\cite{vul4j2022} can be built without human intervention, except for a few old versions of the Spring Framework, since some dependencies cannot be automatically resolved due to the change of their repository URLs\footnote{\url{https://spring.io/blog/2020/10/29/notice-of-permissions-changes-to-repo-spring-io-fall-and-winter-2020}}. To correspond to the common compilation and distribution process for open source libraries, we use the default compiler options and flags specified in the build scripts.

\subsection{Evaluation Metrics}
\label{sec:metrics}

Patch presence test can be considered as a binary classification problem in evaluations, and there are four possible outcomes:

\begin{itemize}
    \item True Positive (\textbf{TP}): The binaries \underline{are} patched, and \appname \underline{detects} the presence of the patch.
    \item True Negative (\textbf{TN}): The binaries \underline{are not} patched, and \appname \underline{does not detect} the presence of the patch.
    \item False Positive (\textbf{FP}): The binaries \underline{are not} patched, but \appname \underline{detects} the presence of the patch.
    \item False Negative (\textbf{FN}): The binaries \underline{are} patched, but \appname \underline{does not detect} the presence of the patch.
\end{itemize}

Based on these possible outcomes, the evaluation metrics we use are defined as follows:

\textbf{Accuracy} is the proportion of correct predictions among the total number of cases examined.
$$\mathrm{Accuracy} = \dfrac{TP + TN}{TP + TN + FP + FN}$$

\textbf{Precision} measures the accuracy in classifying a sample as positive.
$$\mathrm{Precision} = \dfrac{TP}{TP + FP}$$

\textbf{Recall}, also known as True Positive Rate (\textbf{TPR}), measures the ability to detect positive samples.
$$\mathrm{Recall} = \dfrac{TP}{TP + FN}$$

\textbf{F1 Score} is the harmonic mean of precision and recall, which symmetrically represents both precision and recall in one metric.
$$F_1 = 2 \times \dfrac{\mathrm{Precision} \times \mathrm{Recall}}{\mathrm{Precision} + \mathrm{Recall}}$$

\subsection{Implementation of Our Approach}

We implement the framework in Java and the replication package is publicly available.\footnote{\url{https://github.com/pan2013e/ppt4j}} To parse Java source code and analyze ASTs, we exploit Spoon which is a library for implementing analyses and transformations of Java source code proposed by Pawlak et al.~\cite{spoon}.
To parse and analyze Java binaries, we employ ASM~\cite{asm_website}, which is an all-purpose Java bytecode manipulation and analysis framework. We set the threshold of feature equivalence $\sigma_f$ to 0.7 and the threshold of patch support rate $\sigma_p$ to 0.6. Besides, we set the maximum window size of the second feature matching pass to 5 instructions. \appname is lightweight and does not cost too much CPU and memory resources. We perform our experiments on a Java HotSpot(TM) 64-bit server VM (build 17.0.2+8-LTS-86) on a personal computer (MacOS 13, 64-bit 3.2GHz CPU, 16GB RAM).

\section{Results}
\label{sec:eval}

In this paper, we aim to answer the following four research questions:

\begin{enumerate}
\renewcommand{\labelenumi}{\textbf{RQ.\theenumi}}
    \item \textbf{(Effectiveness)} How accurate is the patch presence test framework compared to previous work?
    \item \textbf{(Efficiency)} How efficient is the patch presence test framework, especially when dealing with large code repositories?
    \item \textbf{(Ablation Study)} How do the analyses described in Section~\ref{sec:feat_ext} contribute to the overall effectiveness?
    \item \textbf{(In-the-wild Evaluation)} Can our approach analyze open-source libraries in real-world applications? 
\end{enumerate}

\subsection{RQ1: Effectiveness}
\label{sec:rq1}

\begin{table}[htbp]
\centering
\caption{Test results on the dataset}
\label{tab:res}
\resizebox{0.8\columnwidth}{!}{
\begin{tabular}{cc@{\hspace{1.5em}}rrrr}
\toprule
\multicolumn{2}{c}{\multirow{2}{*}{\textbf{Test Suite}}} &
  \multicolumn{4}{c}{\textbf{Metrics}} \\ \cmidrule(l){3-6} 
\multicolumn{1}{l}{} &
  \multicolumn{1}{l}{} &
  \multicolumn{1}{r}{Acc.} &
  \multicolumn{1}{r}{Prec.} &
  \multicolumn{1}{r}{Recall} &
  \multicolumn{1}{r}{F1} \\ \midrule
\multirow{2}{*}{BScout}   & \dtname{1} & 
                            100\%  & 100\% & 100\%  & 100\%     \\
                          & \dtname{2} & 
                            87.9\% & 100\% & 75.8\% & 86.2\% \\ \midrule
\multirow{2}{*}{\appname} & \dtname{1} & 
                            100\%  & 100\% & 100\%  & 100\%     \\
                          & \dtname{2} & 
                            98.5\% & 100\% & 97.0\% & 98.5\% \\ \bottomrule
\end{tabular}
}
\end{table}

\subsubsection{Results}

The overall result is presented in Table~\ref{tab:res}.
According to Table~\ref{tab:res}, we can observe that both BScout and our work achieve 100\% accuracy, precision, recall and F1 score on \dtname{1}. 
The results of BScout illustrate that our reproduction of it is consistent with the results reported by Dai et al~\cite{dai2020bscout}. 
When we perform experiments on \dtname{2}, which has more subtle patches, the recall and F1 score of BScout drop noticeably, while \appname still maintains a remarkable performance.

In addition to the conclusion mentioned above, \appname does not generate false positive results, which means that it does not mistakenly report a binary as having been patched when it has not. This is an important feature as it ensures that developers can trust the tool's output and avoid wasting time investigating false leads.

\subsubsection{Qualitative Analysis}
We also conduct an qualitative analysis that presents some representative patterns to demonstrate the strengths of our work compared to BScout.

\myding{1} \textbf{Minor changes.} We examine several security patches in our dataset and find that the features extracted by BScout are unable to distinguish some minor changes.

\begin{figure}
\begin{minipage}{\columnwidth}
\centering
\begin{minted}[frame=single, fontsize=\scriptsize, breaklines, linenos, escapeinside=||, numbersep=2pt]{diff}
@@ -174,7 +174,7 @@ public <T> T deserialze(DefaultJSONParser parser, Type type, Object fieldName) {
             componentType = componentClass = clazz.getComponentType();
         }
         JSONArray array = new JSONArray();
-        parser.parseArray(componentClass, array, fieldName);
+        parser.parseArray(componentType, array, fieldName);
 
         return (T) toObjectArray(parser, componentClass, array);
     }
\end{minted}
\captionof{figure}{Patch snippet from CVE-2017-18349~\cite{CVE-2017-18349}: An example of minor change that only replaces an argument}
\label{fig:diff_minor}
\end{minipage}
\begin{minipage}{\columnwidth}
\centering
\includegraphics[width=\columnwidth]{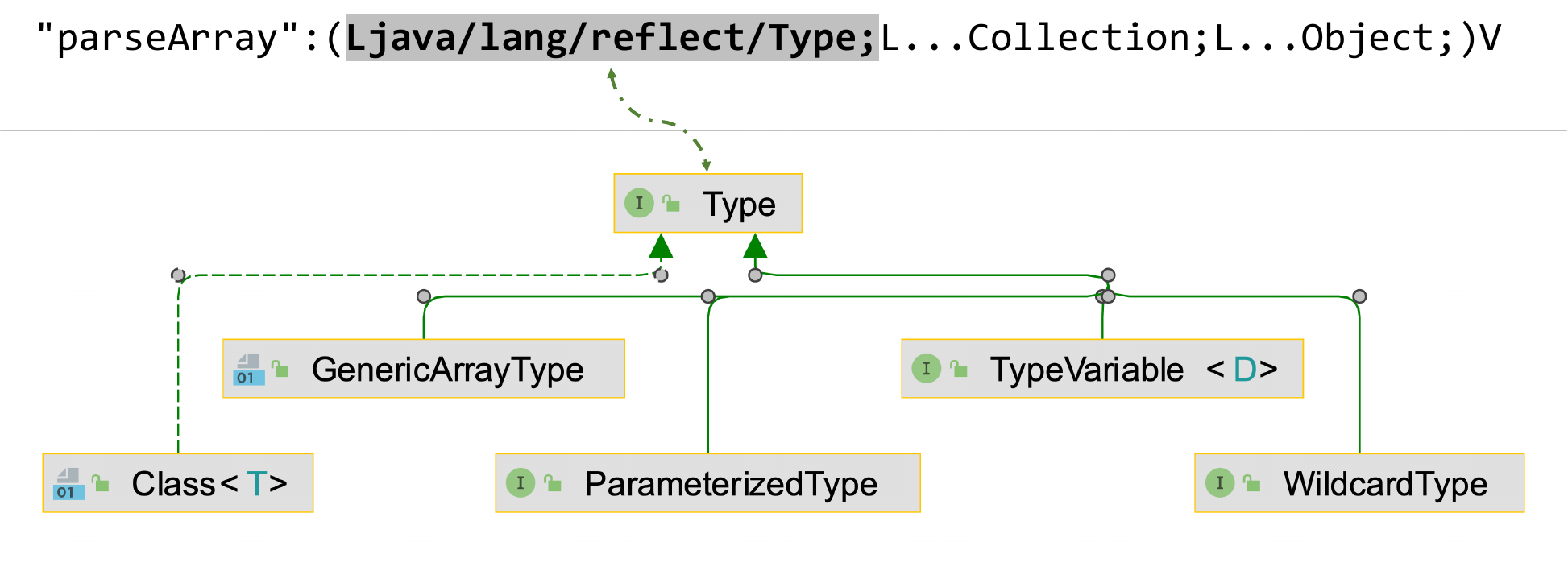}
\Description{Class hierarchy of \texttt{java.lang.reflect.Type}}
\captionof{figure}{Class hierarchy of \texttt{java.lang.reflect.Type} }
\label{fig:type_uml}
\end{minipage}
\Description{Patch snippet from CVE-2017-18349 and Class hierarchy of \texttt{java.lang.reflect.Type}}
\end{figure}

For example, Figure~\ref{fig:diff_minor} shows the patch \diff of CVE-2017-18349. 
In this patch, only the first parameter of the method call \texttt{parseArray\\ (Type, Collection, Object)} is changed from \verb|componentClass| to \texttt{componentType}.
This is a minor change that only includes a variable replacement. 

\faThumbsDown \ \textbf{BScout:} According to Dai et al.~\cite{dai2020bscout},  BScout only extracts method names with argument lengths. In this case, the features of the method call before and after the patch are identical.

\faThumbsUp \ \textbf{\textsc{Ppt4J}:} In this case, \appname works because it implements a type analysis, as described in Section~\ref{sec:feat_ext}. The type of the first parameter in the method signature is \verb|java.lang.reflect.Type|. It is too general in the class hierarchy, and many subclasses implement this interface, as illustrated in Figure~\ref{fig:type_uml}.
With the help of type analysis, \appname traces the variables and checks their exact type instead of using their common super class \texttt{java.lang.reflect.Type}. The exact type of \verb|componentClass| is \verb|java.lang.Class|, and the exact type \verb|componentType| is \verb|java.lang.reflect.Type|. In this way, \appname extracts different features before and after the patch, and thus detects the minor change.

\myding{2} \textbf{Syntactic sugars.} Syntatic sugars~\cite{syntactic_sugar} make modern programming languages carry more semantic information than their lower-level representations such as IR, bytecode, and assembly code. To accurately perform feature matching (as described in Section~\ref{sec:feat_match}), we must bridge the gaps between source code and binary, i.e., interpret syntactic sugars in the source code to bytecode representations. \appname is able to interpret and capture these types of information to reduce the number of false negative results. 

For example, variable-length arguments (abbriviated as \textit{varargs}) are syntactic sugars in Java and are converted into raw arrays in code generation. BScout avoids extracting array creations because Java compilers automatically create \verb|java.lang.Object[]| when calling \textit{vararg} methods~\cite{dai2020bscout}. This solves the inconsistency problem, but BScout cannot detect if a programmer manually creates an array. On the contrary, \appname interprets \textit{varargs} at the source code level, allowing it to detect all kinds of array creations.

\begin{figure}[htbp]
\centering
\begin{minted}[frame=single, fontsize=\footnotesize, breaklines, linenos, escapeinside=||, numbersep=2pt]{diff}
@@ -46,7 +47,7 @@ public abstract class MimeTypeUtils {
    // other codes omitted
 
-   private static final Random RND = new Random();
+   private static final Random RND = new SecureRandom();
\end{minted}
\Description{Patch snippet from CVE-2018-1272}
\caption{Patch snippet from CVE-2018-1272~\cite{CVE-2018-1272}: An example of syntatic sugar that uses a field initialization statement}
\label{fig:diff_field_init}
\end{figure}

Another example of syntactic sugars in Java is field initializations. Field initializations are actually statements in the constructor method \verb|<init>| or static initializing block \verb|<clinit>|. Our work \appname is designed to extract features from field initializations, but BScout simply ignores them. Figure~\ref{fig:diff_field_init} is a security patch of the Spring framework~\cite{spring_framework}, which only changes the initialization of the field \verb|RND|. \appname detects this change, while BScout does not.

\myding{3} \textbf{Semantic redundancy in \textit{diff}.} As shown in Section~\ref{sec:motivation}, \appname utilizes the features of the source code and excludes unrelated information in \textit{diff}. However, BScout takes into account the whole patch \textit{diff}. Although BScout indeed does not miss any semantic information in the patches, it is very likely to introduce unrelated information and fail in cases as shown in Figure~\ref{fig:diff_case3}.

\rqsum{1}{
\appname achieves high accuracy, precision, recall and F1 score on the dataset (98.5\%, 100\%, 97\% and 98.5\% respectively) and outperforms our baseline BScout by 14.2\% in terms of F1 score. In addition, \appname is effective in handling patches with minor changes.
}

\subsection{RQ2: Efficiency}
\label{sec:rq2}

\begin{table}
\centering
\begin{threeparttable}
\caption{Time consumption on the dataset}
\label{tab:perf}
\begin{tabular}{c@{\hspace{2em}}c@{\hspace{2em}}c}
\toprule
\textbf{Framework} & \textbf{Average}  &   $\boldsymbol{\sim}$\textbf{75\%}\tnote{a} \\ \midrule
BScout\tnote{b} &  0.34 sec/patch   &  0.28 sec/patch    \\ \midrule
\appname & 0.48 sec/patch & 
                   0.30 sec/patch      \\ 
\bottomrule
\end{tabular}
\begin{tablenotes}[flushleft]
\small
\item[a] \footnotesize{75\% of test cases are analyzed within this amount of time.}
\item[b] \footnotesize{This refers to our reproduction of Dai et al.'s work~\cite{dai2020bscout}.}
\end{tablenotes}
\end{threeparttable}
\end{table}

To answer RQ2, we measure the time consumption of \appname on the dataset.\footnote{When collecting data, we ignore the startup time of the Java virtual machine and third-party dependencies, and only focus on the components described in Section~\ref{sec:approach}.} Table~\ref{tab:perf} shows that most patches can be quickly analyzed. Some patches may take a bit longer (several seconds) because of the CFG construction and analysis on large bytecodes, but we think it is still acceptable compared to the time costs of human inspections. \appname can achieve this performance because only dependent bytecodes are analyzed, so the time cost is not proportional to the project size. Besides, the preprocessing phase in dataset preparations caches the source code features and reduces the analysis time when users input their binaries. Table~\ref{tab:perf} also lists the time consumption of our reimplemented baseline, BScout. Compared to BScout, \appname is a bit slower in most test cases, but we think that \appname's advantages in effectiveness (as discussed in RQ1) can compensate for the shortcomings in terms of time efficiency.

\rqsum{2}{
\appname analyzes most security patches in one second and is also fast when analyzing binaries from large projects.
}

\subsection{RQ3: Ablation Study}

To investigate the contribution of an analysis and its corresponding feature type (as described in Section~\ref{sec:feat_ext}) to the effectiveness of \appname, we create the following variants:

\begin{itemize}
    \item \textsc{Ppt4J}\underline{~~}\verb|FULL|: The complete version of \appname, exactly as Section~\ref{sec:approach} illustrated.
    \item \appvar{1}: Type analysis is removed. This means that method signatures are not revised during feature extraction.
    \item \appvar{2}: Some special instructions are ignored, such as loop and branch.
    \item \appvar{3}: Constant propagation/folding is removed. This means that constant fields and expressions are not simplified during feature extraction.
\end{itemize}

\begin{figure}
    \centering
    \includegraphics[width=\columnwidth]{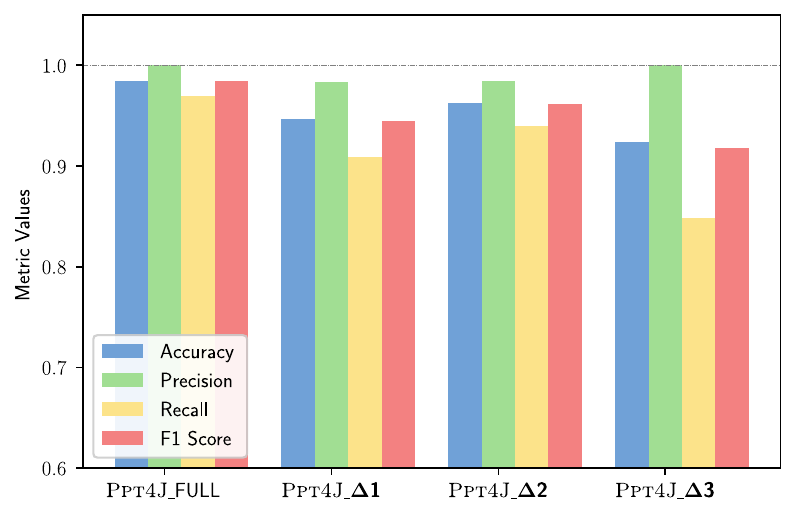}
    \Description{Test results for different variants of \textsc{Ppt4J}}
    \caption[]{Test results for different variants of \textsc{Ppt4J}}
    \label{fig:abl}
\end{figure}

We evaluate the above four variants on the same dataset (as described in Section~\ref{sec:dataset}), and compare the incomplete ones with \textsc{Ppt4J}\underline{~~}\verb|FULL|. The results are illustrated in Figure~\ref{fig:abl}. We can conclude that \appname benefits from these analyses. In the cases of \appvar{1-3}, the F1 score decreases by 4.0\%, 2.3\%, and 6.8\% respectively after one specific analysis is removed. 

\rqsum{3}{The analyses proposed in Section~\ref{sec:feat_ext} make extracted features fine-grained, and they indeed improve the effectiveness of \appname.}

\subsection{RQ4: In-the-wild Evaluation}

\subsubsection{Experimental setting}
To evaluate the effectiveness of our approach in practical settings, we use IntelliJ IDEA Ultimate~\cite{intellij_idea} as the target software for our in-the-wild evaluation. The reason to choose IntelliJ IDEA is that it is a widely used IDE software, and it embeds a large number of binaries from open-source Java libraries. We obtain various versions of IntelliJ IDEA and extract the included third-party libraries. Next, we conduct patch presence tests on these open-source libraries for specific patches. However, labeling real-world binaries with ground truth is challenging since not all binary packages contain version information. To address this issue, we utilize unit tests in some patches. These unit tests are committed together with the bug fixes, and can be trusted sources of ground truths. If our patch presence predictions align with the unit test results, it indicates that our approach can perform well in real-world scenarios. 
Specifically, each test case employed in this experiment includes:
\myding{1} Binaries of a third-party library extracted from IntelliJ IDEA.
\myding{2} A security patch, along with the unit test, and the source code of the library before and after the patch.

\begin{table}
\centering
\caption{Evaluation on different versions of IntelliJ IDEA}
\label{tab:idea}
\resizebox{\columnwidth}{!}{
\begin{threeparttable}[htbp]
\setlength{\tabcolsep}{7pt} 
\renewcommand{\arraystretch}{1.5}
\begin{tabular}{@{}lcclclclcl@{}}
\toprule
\multicolumn{1}{c}{} &
  \multicolumn{9}{c}{\textbf{Version Timeline}\tnote{a}} \\ \cmidrule(l){2-10} 
\multicolumn{1}{c}{\multirow{-2}{*}{\textbf{Vulnerability}}} &
  \texttt{V1} &
  \multicolumn{2}{c}{\texttt{V2}} &
  \multicolumn{2}{c}{\texttt{V3}} &
  \multicolumn{2}{c}{\texttt{V4}} &
  \multicolumn{2}{c}{\texttt{V5}} \\ \midrule
CVE-2019-12402 \faCodeBranch\texttt{:08/19}\tnote{b} &
  {\color[HTML]{343434} TN} &
  \multicolumn{2}{c}{{\color[HTML]{343434} TN}} &
  \multicolumn{2}{c}{\cellcolor[HTML]{A9F7A8}{\color[HTML]{343434} TP}} &
  \multicolumn{2}{c}{\cellcolor[HTML]{A9F7A8}{\color[HTML]{343434} TP}} &
  \multicolumn{2}{c}{\cellcolor[HTML]{A9F7A8}{\color[HTML]{343434} TP}} \\
CVE-Anonymous-1 &
  {\color[HTML]{343434} TN} &
  \multicolumn{2}{c}{{\color[HTML]{343434} TN}} &
  \multicolumn{2}{c}{\cellcolor[HTML]{FFCCC9}{\color[HTML]{343434} TN}} &
  \multicolumn{2}{c}{\cellcolor[HTML]{FFCCC9}{\color[HTML]{343434} TN}} &
  \multicolumn{2}{c}{\cellcolor[HTML]{FFCCC9}{\color[HTML]{343434} TN}} \\
CVE-Anonymous-2\tnote{c} &
  {\color[HTML]{343434} TN} &
  \multicolumn{2}{c}{{\color[HTML]{343434} TN}} &
  \multicolumn{2}{c}{\cellcolor[HTML]{FFCCC9}{\color[HTML]{343434} TN}} &
  \multicolumn{2}{c}{\cellcolor[HTML]{FFCCC9}{\color[HTML]{343434} TN}} &
  \multicolumn{2}{c}{\cellcolor[HTML]{FFCCC9}{\color[HTML]{343434} TN}} \\
CVE-2021-29425 \faCodeBranch\texttt{:05/18} &
  {\color[HTML]{343434} TN} &
  \multicolumn{2}{c}{\cellcolor[HTML]{FFCCC9}{\color[HTML]{343434} TN}} &
  \multicolumn{2}{c}{\cellcolor[HTML]{A9F7A8}{\color[HTML]{343434} TP}} &
  \multicolumn{2}{c}{\cellcolor[HTML]{A9F7A8}{\color[HTML]{343434} TP}} &
  \multicolumn{2}{c}{\cellcolor[HTML]{A9F7A8}{\color[HTML]{343434} TP}} \\
HTTPCLIENT-1803 \faCodeBranch\texttt{:01/17} &
  \cellcolor[HTML]{FFCCC9}{\color[HTML]{343434} TN} &
  \multicolumn{2}{c}{\cellcolor[HTML]{FCE38A}{\color[HTML]{343434} FN}} &
  \multicolumn{2}{c}{\cellcolor[HTML]{FCE38A}{\color[HTML]{343434} FN}} &
  \multicolumn{2}{c}{\cellcolor[HTML]{FCE38A}{\color[HTML]{343434} FN}} &
  \multicolumn{2}{c}{\cellcolor[HTML]{FCE38A}{\color[HTML]{343434} FN}} \\
CVE-2017-1000487 \faCodeBranch\texttt{:10/13} &
  \cellcolor[HTML]{A9F7A8}{\color[HTML]{343434} TP} &
  \multicolumn{2}{c}{\cellcolor[HTML]{A9F7A8}{\color[HTML]{343434} TP}} &
  \multicolumn{2}{c}{\cellcolor[HTML]{A9F7A8}{\color[HTML]{343434} TP}} &
  \multicolumn{2}{c}{\cellcolor[HTML]{A9F7A8}{\color[HTML]{343434} TP}} &
  \multicolumn{2}{c}{\cellcolor[HTML]{A9F7A8}{\color[HTML]{343434} TP}} \\
CVE-2015-6748 \faCodeBranch\texttt{:07/15} &
  {\color[HTML]{343434} N/A\tnote{d}} &
  \multicolumn{2}{c}{\cellcolor[HTML]{A9F7A8}{\color[HTML]{343434} TP}} &
  \multicolumn{2}{c}{\cellcolor[HTML]{A9F7A8}{\color[HTML]{343434} TP}} &
  \multicolumn{2}{c}{\cellcolor[HTML]{A9F7A8}{\color[HTML]{343434} TP}} &
  \multicolumn{2}{c}{\cellcolor[HTML]{A9F7A8}{\color[HTML]{343434} TP}} \\
CVE-2015-6420 \faCodeBranch\texttt{:11/15} &
  \cellcolor[HTML]{FFCCC9}{\color[HTML]{343434} TN} &
  \multicolumn{2}{c}{\cellcolor[HTML]{A9F7A8}{\color[HTML]{343434} TP}} &
  \multicolumn{2}{c}{\cellcolor[HTML]{A9F7A8}{\color[HTML]{343434} TP}} &
  \multicolumn{2}{c}{\cellcolor[HTML]{A9F7A8}{\color[HTML]{343434} TP}} &
  \multicolumn{2}{c}{\cellcolor[HTML]{A9F7A8}{\color[HTML]{343434} TP}} \\ \bottomrule
\end{tabular}
\renewcommand{\arraystretch}{1}
\begin{tablenotes}[flushleft]
\small
\item [a] \footnotesize{\texttt{V1} - \texttt{V5} are 5 versions of IntelliJ IDEA Ultimate, sorted in ascending order of release time. V1: IU-181.5684.4; V2: IU-191.8026.42; V3: IU-203.8084.24; V4: IU-213.7172.25; V5: IU-231.8109.175. The first two digits in the version string specify the release year, e.g., V1 was released in 2018.
}
\item [b] \footnotesize{Patch commit time. Retrieved from Github, in MM/YY format.}
\item [c] \footnotesize{Real IDs of \textit{CVE-Anonymous-1/2} are omitted due to ``responsible reporting'' principle. }
\item [d] \footnotesize{N/A means this version of software does not include the library.}
\end{tablenotes}
\end{threeparttable}
}
\end{table}

\subsubsection{Results}
The detailed result is shown in Table~\ref{tab:idea}. We conclude that most of the output from \appname is consistent with the unit tests and the accuracy is 89.7\%. We also notice that \appname does not generate false positive outputs. In conclusion, we believe that \appname is capable of real world scenarios.
We also learn some facts about the patch status in IntelliJ IDEA from the experiment results.

\myding{1} The application vendor JetBrains promptly applied the upstream patch for CVE-2019-12402, ensuring that libraries in releases after 2020 were not affected by this vulnerability. However, even though the patch commit for CVE-2021-29425 was released in May 2018, JetBrains did not apply it in one of their 2019 versions, such as \textit{IU-191.8026.42}. Similarly, for CVE-2015-6420, the patch commit was available as early as November 2015, but they failed to apply it in one of the 2018 versions.

\myding{2} We notice that a third-party library in IntelliJ IDEA has not yet been patched for two specific vulnerabilities (i.e., \textit{CVE-Anonymous-1} and \textit{CVE-Anonymous-2}, as shown in Table~\ref{tab:idea}) until now. The application vendor JetBrains forked its branch, but has not merged the upstream branch later, leaving the vulnerabilities unresolved. 
Although we cannot definitively assert that vulnerabilities in open source libraries will invariably impact commercial software, we believe that minimizing vulnerabilities in these libraries can mitigate the risk of exploitation. Thus, we report this problem to JetBrains. 

\myding{3} We also perform this evaluation using our reimplemented baseline. Compared to \appname's results, the accuracy of BScout decreases by 14.3\% and drops to 76.9\%. Upon examination of the failed test cases, we find out \appname performs better due to its effectiveness in handling patches with minor changes, as discussed in RQ1. For example, in the case of \textit{CVE-Anonymous-1}, BScout cannot even tell whether a binary contains the patch because no features are extracted from the patch.

\rqsum{4}{
\appname is capable of analyzing open source libraries in real-world applications. In our evaluation, it achieves an accuracy of 89.7\% and detects two unpatched CVEs in a third-party library within IntelliJ IDEA. We have reported this potential problem to JetBrains.
}

\section{Discussion}
\label{sec:disc}

\subsection{Line number information}
\label{sec:line_number}

Although generated by default, line number information is part of the debug information, and can be stripped from binaries. \appname fails without such information. Despite this, we perform an empirical study on line number information in open-source Java libraries. We collect 9,077 jar files (containing 2,032,221 class files in total), which are downloaded from the Maven central repository~\cite{maven}. Our findings show that over 90\% of these class files include line number information. As \appname is proposed for use for general open-source libraries, we believe it is still applicable in many practical scenarios.

\subsection{\textsc{Ppt4J} vs. Baseline}
Compared to the baseline BScout, \appname selects different sets of feature types, and applies different algorithms to extract non-trivial features and handle syntactic sugars. With these algorithms, \appname is able to capture minor changes in patches and the extracted features are more fine-grained, as illustrated in Section~\ref{sec:feat_ext} and the qualitative analysis on the dataset in Section~\ref{sec:rq1}. It is also worth noting that we propose the semantic change generator in \appname. With semantic change generation, \appname is able to process \diff files and filter out unrelated information based on the previously extracted features, as illustrated in Section~\ref{sec:bg} and Section~\ref{sec:approach}. However, BScout takes the complete patch \diff as input and cannot handle the semantic redundancy in diffs. The drawback of \appname compared to BScout is its efficiency and dependence on line number information, but as discussed earlier in Section~\ref{sec:rq2} and Section~\ref{sec:line_number}, we believe \appname remains practical.

\section{Threats to Validity}
\label{sec:discussion}

\noindent\textbf{Correctness of our reproduction.} 
\myding{1} As mentioned earlier in Section~\ref{sec:dataset} and Section~\ref{sec:rq1}, we reimplement BScout~\cite{dai2020bscout} and reproduce identical results on dataset \dtname{1} in terms of accuracy, precision, recall and F1 score. However, in addition to \dtname{1}, Dai et al.~\cite{dai2020bscout} also construct a dataset consisting of Android applications and evaluate their approach on this dataset. Since the details of this dataset are not available, we cannot reproduce all experiments in the authors' paper. This might be a threat to the correctness of our reproduction and the effectiveness evaluation. \myding{2}  We only reimplement one variant of BScout that relies on line number information. The complete version of BScout leverages a machine learning model, but neither the model weights nor the training dataset is publicly available. Thus, it is not practical for us to reimplement the complete one. However, this does not create issues with the reliability of our findings in Section~\ref{sec:eval}. According to Dai et al.~\cite{dai2020bscout}, the performance of the complete version of BScout is lower than the variant if line number information is present. Therefore, the performance of our reimplemented baseline reported in Section~\ref{sec:eval} is supposed to be an upper bound.

\noindent\textbf{Patch backporting.} 
Backporting security patches to older versions is a common practice, which means that multiple different patch commits can correspond to the same vulnerability. However, when constructing the dataset, we only selected one version for each vulnerability, and \appname cannot automatically select the correct patch commit. This limitation can lead to erroneous results when \appname analyzes open-source libraries in real-world scenarios.

\noindent\textbf{Inherent flaws with rule-based features.} 
\appname extracts rule-based features from the source code, providing finer granularity than the baseline approach. However, this method has an inherent limitation in that it cannot always cover all language features. As the Java language continues to evolve, the tool must be updated to adapt to programs written using new standards.

\noindent\textbf{Effectiveness against version changes.} 
In \appname, we use the longest common subsequence algorithm to match the features. However, this algorithm assumes that the source code structure is stable and may produce unsatisfactory results in the case of significant code changes, such as API-breaking updates.

For example, during our evaluation of IntelliJ IDEA, we encounter four false negative samples related to \textit{HTTPCLIENT-1803} (in Table~\ref{tab:idea}). In this case, the unit tests show that the samples were patched, but \appname fails to detect the patch.

Upon further examination of the bytecodes, we discover that IntelliJ IDEA imported a library of a newer version, which has been reconstructed and replaced by a different set of APIs. As a result, the code differences in the security patch were not reflected in the binaries of the new version.

\section{Related Work}
\label{sec:related}

\noindent\textbf{Binary similarity analysis.} 
The study of function similarity in binaries has been tackled through various techniques. Bourquin et al.~\cite{bourquin2013binslayer} and BinDiff~\cite{bindiff} rely on the isomorphism of control flow graphs to evaluate similarities.
Khoo et al. propose Rendezvous~\cite{khoo2013rendezvous}, which optimizes this approach by improving the granularity of the analysis. 
Alternative approaches, such as BinHunt~\cite{gao2008binhunt} and iBinHunt~\cite{ming2013ibinhunt}, formulate the semantic equivalence of basic blocks through symbolic execution and theorem provers. 

For cross-platform capability, Pewny et al.~\cite{pewny2015multimh} extract I/O behaviors at the basic block level. In addition, Eschweiler et al. propose discovRE~\cite{eschweiler2016discovre}, which generates platform-independent feature vectors from basic blocks. 
Furthermore, new techniques have been proposed to address the issues of efficiency and scalability. Genius~\cite{feng2016genius} encodes the representations of control flow graphs as graph embeddings. Other methods like Gemini~\cite{xu2017gemini} further utilize neural networks. Additionally, Huang et al.~\cite{huang2017binsequence} propose BinSequence, which uses Min-hashing to filter the search space.

\noindent\textbf{Patch presence test.} 
The first to publicly propose and implement patch presence tests is FIBER~\cite{fiber}. FIBER generates binary signatures by analyzing security patches. These signatures reflect representative changes introduced by the patches and are utilized to search for the target binary file.

Subsequent papers improve and extend FIBER's approach. Dai et al. propose BScout~\cite{dai2020bscout} to check the existence of patches in Java binaries without generating signatures. BScout employs new techniques to bridge the gap between source code and bytecode instructions and to check patch semantics throughout the target executable file. Jiang and Zhang propose PDiff~\cite{pdiff} to test the patch presence of images from the downstream kernel in the open source kernel domain. PDiff generates summaries for patches. Based on semantic summaries, PDiff compares the target with mainstream versions before and after applying the patch and selects the closest for evaluation. An alternative approach, Osprey~\cite{osprey}, employs a more lightweight static analysis algorithm compared to FIBER and improves efficiency without compromising too much accuracy. Among the related work, FIBER, PDiff and Osprey test the patch presence of C/C++ binaries, while BScout and our proposed \appname target Java binaries. Compared to the state of the art, the advantages of \appname are mainly its superiority of capturing minor changes in patches and its ability to extract precise semantic changes from patch diffs.
\section{Conclusion and Future Work}
\label{sec:conclusion}

In this paper, we aim to address the problem with patch \textit{diffs} in existing work by highlighting semantic changes. Then we design and implement \appname, a patch presence test framework targeting Java binaries, which performs accurate tests with the help of semantic changes that reflect differences in program effects. \appname is systematically evaluated on a dataset with real-world vulnerabilities. The results show that \appname achieves an F1 score of 98.5\% while maintaining a reasonable performance. In addition, we perform an in-the-wild evaluation \appname on IntelliJ IDEA, in which it achieves an accuracy of 89.7\% with no false positive results. \appname detects two unpatched CVEs in a third-party library within IntelliJ IDEA, and we have reported this potential problem to the vendor. 
In future work, one possible research attempt is to minimize the impact of significant code changes in binaries (e.g., patch backporting and version changes, as described in Section~\ref{sec:discussion}). This investigation may help improve the effectiveness of our proposed approach in real-world applications.
\begin{acks}
This research is supported by the Fundamental Research Funds for the Central Universities (No. 226-2022-00064), National Natural Science Foundation of China (No. 62141222), and the National Research Foundation, under its Investigatorship Grant (NRF-NRFI08-2022-0002). Any opinions, findings and conclusions or recommendations expressed in this material are those of the author(s) and do not reflect the views of National Research Foundation, Singapore.
\end{acks}

\bibliographystyle{ACM-Reference-Format}
\bibliography{ref}

\end{document}